\documentclass[prd,twocolumn,showpacs,showkeys,amsmath,amssymb]{revtex4-1}

\usepackage{graphicx}
\usepackage[version=3]{mhchem}
\usepackage{hepnames}

\input{definitions}

\begin{document}

\markboth{Michal Kreps}
{Measurement of \CP violation in \BsJpsiphi decay}

%%%%%%%%%%%%%%%%%%%%% Publisher's Area please ignore %%%%%%%%%%%%%%
%\catchline{}{}{}{}{}
%%%%%%%%%%%%%%%%%%%%%%%%%%%%%%%%%%%%%%%%%%%%%%%%%%%%%%%%%%%%%%%%%%%

\title{
Measurement of \CP violation in \BsJpsiphi decay 
}

\author{\footnotesize MICHAL KREPS}

\affiliation{Physics Department, University of Warwick\\ Gibbet
Hill Road, Coventry, CV4 7AL, United Kingdom \\
M.Kreps@warwick.ac.uk}


%\pub{Received (Day Month Year)}{Revised (Day Month Year)}

\begin{abstract}
We briefly discuss measurements of \CP violation in \BsJpsiphi decay. Both the phenomenology of 
\Bsboth mixing and the importance of the measurement to searches for new physics, as well as technical 
details and issues with the analysis are included. While current results are consistent with the standard 
model, even large contributions from new physics cannot be excluded.

\end{abstract}

\maketitle

\section{Introduction}

$b$-physics dates back to 1964 when the 
decay of the long lived kaon to two pions, and thus \CP
violation was observed \cite{Christenson:1964fg}. It did not
took very long until a proposal for the theoretical explanation of
\CP violation was made. In their famous work, Kobayashi
and Maskawa showed that with 4 quarks there is no reasonable
way to include \CP violation \cite{Kobayashi:1973fv}.
Together with this, they also proposed several models to
explain the observed \CP violation in the kaon system, amongst which the
six quark model became favored over time.

The explanation of \CP violation in the six-quark model
of Kobayashi and Maskawa builds on the idea
of quark mixing introduced by Cabibbo. The quark mixing
introduces a difference between eigenstates of the strong and weak
interactions.  \CP violation
requires a complex phase in order to provide a difference
between a process and its charge conjugate.
In the four-quark model, the quark mixing is described
by a $2\times2$ unitarity matrix. With only four quarks, states
can be always re-phased in order to keep the mixing matrix real
and thus quark mixing cannot accommodate the observed \CP violation.
With the extension to six quarks, the mixing
matrix becomes a $3\times3$ unitarity matrix, called
the Cabibbo-Kobayashi-Maskawa matrix, $V_{CKM}$. In this case there is
no possibility to rotate away all phases and one complex
phase always remains in the matrix. This complex phase of
$V_{CKM}$ provides the \CP
violation in the standard model. This idea had two important
implications. First, in addition to the three quarks known in
the early 1970's and the predicted charm quark, it postulates the 
existence of two additional quarks, called bottom and top.
Second, despite the tiny \CP violation in the kaon system,
the proposed mechanism implies large \CP violation in the \Bzero
system. It took almost three decades, but both predictions
were experimentally confirmed, first by the discovery of the bottom
quark in 1977 \cite{Herb:1977ek} followed by that of the top quark
in 1995\cite{Abachi:1994td,Abe:1995hr} and
finally by the measurement of large \CP violation in the \Bzero
system in 2001 \cite{Aubert:2001sp,Abe:2001xe}. The observation of 
large \CP violation in \Bzero decays confirmed the Kobayashi-Maskawa mechanism 
as the way to generate the \CP violation in the standard model, this resulting
in the 2008 Nobel prize for Kobayashi and Maskawa.

After confirmation of the standard model, the focus shifted to the search for new physics. 
One of the most promising processes is \Bs mixing, governed by the CKM
matrix element $V_{ts}$.  Indirect information suggests
$V_{ts}$ to be almost real, which translates to the fact that
the \CP violation due to the \Bs mixing is expected to be tiny 
in the standard model. Amongst possible ways of accessing new physics in the \Bsboth mixing,
the measurement of \CP violation in \BsJpsiphi decay is considered as
a golden way. In this paper we review the current status of the existing measurements
in the \BsJpsiphi decay, as well as the issues connected to the measurement itself. 
We will omit here many details in favour of providing
a comparison of measurements from different experiments and
discussing features which are not always emphasised. Details
of separate measurements are available for CDF in 
Refs.~\cite{Aaltonen:2007he,Aaltonen:2008hg,Aaltonen:2010hg}, 
for D\O{} in Refs.~\cite{Abazov:2008fj,Abazov:2010hg} and for LHCb 
in Ref.~\cite{Aaij:2011gd}.

This paper is organized as follows: In section
\ref{sec:phenomenology}, we briefly discuss the phenomenology of
\Bs oscillations. An overview of the basic components of
the analysis is given in section \ref{sec:overview}. In
sections \ref{sec:tagging} and \ref{sec:fit}, we discuss
in more detail, two main components, namely the determination of whether \Bs was produced as
\Bs or \Bsbar and the maximum
likelihood fit.  Section \ref{sec:statistics} deals with
the statistical issues encountered in the measurements. The results
are given in section \ref{sec:results} and the paper concludes
with future prospects in section \ref{sec:prospects}.

Note on particle naming. As this topic requires the distinction between the use of both particle and
anti-particle and the relevant single-flavour cases, we adopt a notation in which \Bsboth denotes both 
particle \Bs and anti-particle \Bsbar, while the use of \Bs or \Bsbar denotes a given flavour.
\vfill

\section{Phenomenology of the \Bs system}
\label{sec:phenomenology}

The time evolution of the \Bs system is described by the
Schr\"odinger equation
\begin{displaymath}
i \frac{d}{dt}
\left(
\begin{array}{c}
| \Bs(t) \rangle \\ | \Bsbar (t) \rangle
\end{array}
\right)
=
\left( \hat{M} - \frac{i}{2} \hat{\Gamma} \right)
\left(
\begin{array}{c}
| \Bs(t) \rangle \\ | \Bsbar (t) \rangle
\end{array}
\right),
\end{displaymath}
where the $2\times 2$ matrices $\hat{M}$ and $\hat{\Gamma}$ describe the masses and decay
rates. Diagonalization leads to eigenstates with definite masses and lifetimes
\begin{eqnarray}
|\BsH\rangle &=&p \; |\Bs\rangle + q \; |\Bsbar\rangle, \nonumber \\
|\BsL\rangle &=&p \; |\Bs\rangle - q \; |\Bsbar\rangle, \nonumber 
\end{eqnarray}
with $p$ and $q$ being complex numbers satisfying $|p|^{2} +
|q|^{2} =1$. The states $|\BsH\rangle$ and
$|\BsL\rangle$ have distinct masses $M_\mathrm{H}$ and $M_\mathrm{L}$
and distinct decay widths $\Gamma_\mathrm{H}$ and $\Gamma_\mathrm{L}$.
The \Bs mixing Feynman diagrams responsible for the transition of \Bs to
\Bsbar and vice versa give rise to the off-diagonal
elements $M_{12}$ and $\Gamma_{12}$. The \Bsboth mixing
observables are defined to be the mass difference
between $|\BsH\rangle$ and $|\BsL\rangle$ states
\begin{equation}
 \Delta M =
M_\mathrm{H} - M_\mathrm{L}  = 
        2 { |M_{12}|} %\left( 1 +
\end{equation}
and the decay width difference
\begin{equation}
\Delta\Gamma = \Gamma_\mathrm{L} - \Gamma_\mathrm{H}  = 
        2  |\Gamma_{12}| \cos  \phi_s, 
\end{equation}
where $\phi_s=\mbox{arg}( -M_{12}/\Gamma_{12})$. The standard model
predictions \cite{Lenz:2006hd,Lenz:2011ti} are $\Delta
M=(17.3\pm 2.6)$ \invps, $\phi_s=(0.22\pm0.06)^\circ$ and $\Delta\Gamma=(0.087\pm0.021)$ \invps. 
Physics beyond the standard model can alter the picture by affecting both $M_{12}$ and $\Gamma_{12}$.
In practice most of the models of new physics consider only changes in $|M_{12}|$ and $\phi_s$, and
leave $|\Gamma_{12}|$ unaffected.  The \Bs mixing frequency $\Delta M$ is
measured most precisely by the CDF experiment \cite{Abulencia:2006ze} as $\Delta M=17.77\pm0.10\pm0.07$ 
\invps, and preliminary with similar precision
also by LHCb \cite{Aaij:2011kl} as $\Delta M=17.63\pm0.11\pm0.04$ \invps.
Since $\Delta M$ is precisely known and consistent
with the standard model expectation, new contributions to $|M_{12}|$ are 
strongly constrained. At the same time, $\Gamma_{12}$
is dominated by the tree level $b\rightarrow c\bar c s$
transition and therefore it is non-trivial to construct a model which would
affect $|\Gamma_{12}|$ significantly while at the same time avoiding 
constraints from existing measurements. The phase $\phi_s$ was experimentally 
unconstrained until recently, which made it the prime candidate for searches 
for new physics.

The \Bsboth mixing phase $\phi_s$ can be accessed experimentally
by measurements of the \CP asymmetry in flavour-specific decays or by \CP violation due to an interference
between decays with and without mixing. The flavour-specific \CP asymmetry is defined as
\begin{equation}
a_{fs}^s = {  \frac{|\Gamma_{12}|}{|M_{12}|}} 
 {\sin \left( { \phi_s}\right)}.
\end{equation}
The challenge of the $a_{fs}$ measurement is in the smallness of the effect. If we assume no new physics
contribution to $|\Gamma_{12}|$, then the maximum effect would be typically at most $5\times 10^{-3}$
and would not reach values above $10^{-2}$ even in the most optimistic scenario.

In the second type of measurement, one exploits a final state which is common to both \Bs and \Bsbar.
In such a case an interference between the amplitude for a direct decay of \Bs to a given final state and
the amplitude for \Bs oscillating to \Bsbar which then decays to the final state gives rise to a 
time-dependent \CP violation. The decay \BsJpsiphi which we are going to discuss in this review belongs to this
category. A complication arises from the fact that this type of \CP violation does not measure
directly the $\phi_s$ but rather the relative phase between $\phi_s$ and the phase of
the decay. For the decay \BsJpsiphi the standard model expectation \cite{Charles:2004jd,Bona:2006ah} 
for the \CP-violating phase is
\begin{equation}
\beta_s^\mathrm{SM}=\arg\left(\frac{-V_{ts}V_{tb}^{*}}{V_{cs}V_{cb}^{*}}\right)\approx0.02.  % Give reference
\end{equation}
As mentioned before, the current focus is on the search for new physics. If new physics contributes to the
\Bsboth mixing process, it can modify the phase between $M_{12}$ and $\Gamma_{12}$ to
$\phi_s=\phi_s^\mathrm{SM}+\phi_s^\mathrm{NP}$ where $\phi_s^\mathrm{NP}$ is a new physics contribution. 
This new phase will then also be reflected in the measurement of \BsJpsiphi decay where the observable phase 
can be expressed as 
\begin{equation}
\phiS=-2\beta_s^\mathrm{SM}+\phi_s^\mathrm{NP}.
\end{equation}
Here we have neglected possible higher order corrections to the decay both in the standard model as
well as in new physics models. With the current generation of measurements the sensitivity is not
sufficient to reach standard model values, but it is possible to search for large new physics
contributions. 
While precise predictions for new physics models are not easily available, large effects are
possible even after taking into account all existing constraints on possible new physics
models. It is also worth noting that in many models of new physics, there is a correlation between  
the \CP violation in \Bsboth mixing and other flavour physics observables which can help to
distinguish different new physics scenarios. For examples of a new physics
discussion see Refs.~\cite{Hou:2006mx,Buras:2010pm,Lenz:2010gu,Buras:2010mh,Bauer:2009cf,Soni:2010xh,Chiang:2009ev,Chen:2009bj}.

Finally one remark on the notation which differs between experiments. Different symbols are used to
denote the \CP violation in \BsJpsiphi decay and it is not always made fully clear how they are
defined. On the other hand, given the precision of the existing measurements, the standard model
contributions to the various phases can be neglected and we can interpret results as constraints on new
physics contributions to the \Bsboth mixing phase $\phi_s$. In the following we will use the symbol \phiS to
denote the phase measured by experiment in \BsJpsiphi decay. This relates to $\beta_s$ used by CDF as
$\phiS=-2\beta_s$ and is the same as the $\phi_s$ used by LHCb. In their latest documents, D0 uses the symbol
\phiS.

\section{Measurement overview}
\label{sec:overview}

The measurement of \CP violation in \BsJpsiphi decays is based on a time-dependent analysis. As
the $\Jpsi\phi$ final state is common to both \Bs and \Bsbar, direct decays of the \Bs to $\Jpsi\phi$ are 
possible, as well as the decay where the \Bs first oscillates to \Bsbar and then decays. The two 
paths to the final state interfere and give rise to the \CP violation. The main principle is to extract the 
time-dependent asymmetry
\begin{equation}
  A(t)=\frac{N(\Bs,t)-N(\Bsbar,t)}{N(\Bs,t)+N(\Bsbar,t)},
\end{equation}
where $N(\Bs,t)$ ($N(\Bsbar,t)$) is the number of \BsJpsiphi decays at a given time $t$ where \Bs
(\Bsbar) was originally produced. Using this we can identify the main components of the analysis, which
are the reconstruction and selection of the candidates, the measurement of the decay time for each candidate, 
and finding out whether a given candidate was produced as \Bs or \Bsbar. While the concept is rather simple,
the fact that both \Jpsi and $\phi$ are spin--1 particles means that the final state is a mixture of
\CP-eigenstates. This adds to the complexity of the task as one needs to separate also \CP-odd and
\CP-even components through an angular analysis. Therefore in the final step, rather than
forming an asymmetry, a maximum likelihood fit is typically used to extract information on the \CP
violation. The first two steps are rather straightforward and we will discuss them shortly in this
section; the others are those which require non-trivial work and therefore we discuss them in dedicated
sections.

% Selection
In all three experiments the \BsJpsiphi decays are reconstructed using the \Jpsi decays to two
oppositely charged muons and the $\phi$ decays to oppositely charged kaons. 
The selection uses similar inputs in all three experiments. Events are typically selected by placing
requirements on the momenta of reconstructed \Bsboth candidates and their daughters, the quality of tracks, 
and the quality of kinematic fits where all four tracks are constrained to originate from a common 
vertex. LHCb, and to some extent also CDF, have particle identification capabilities, which are used to
separate kaons from more abundant pions. While the LHCb and D0 experiments use sequential
rectangular requirements, the CDF selection employs a neural network to distinguish signal events from
a combinatorial background. An important aspect of the selection at CDF is that the requirement on the
resulting neural network output is chosen to minimize the expected uncertainties on the measured \CP
violation. This is achieved by performing the analysis on simulated experiments, where each simulated
experiment uses the number of signal and background events corresponding to the given requirement 
on the neural network output.
\begin{figure*}[pht]
\begin{center}
\includegraphics[width=3.0in]{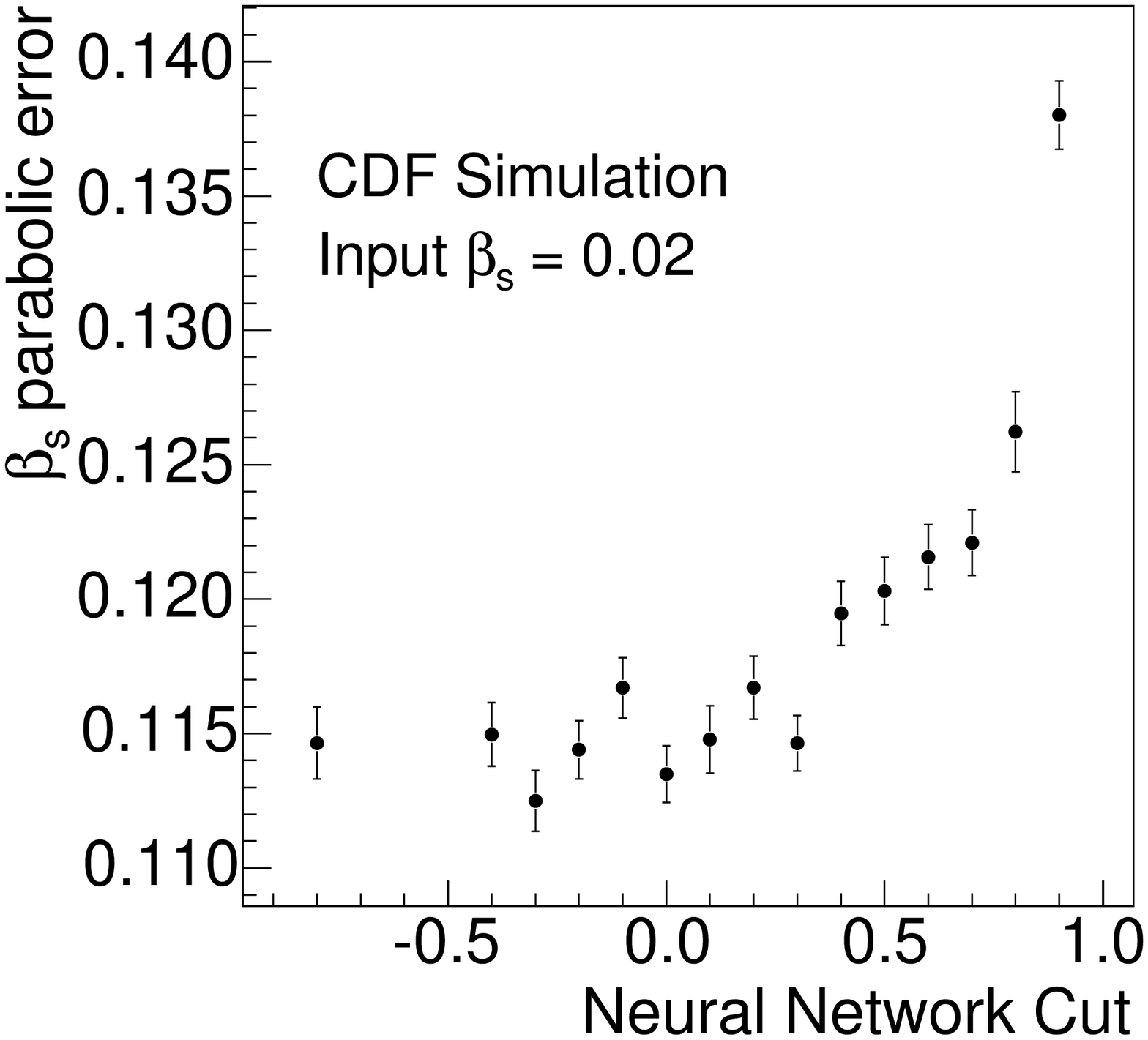}
\includegraphics[width=3.5in]{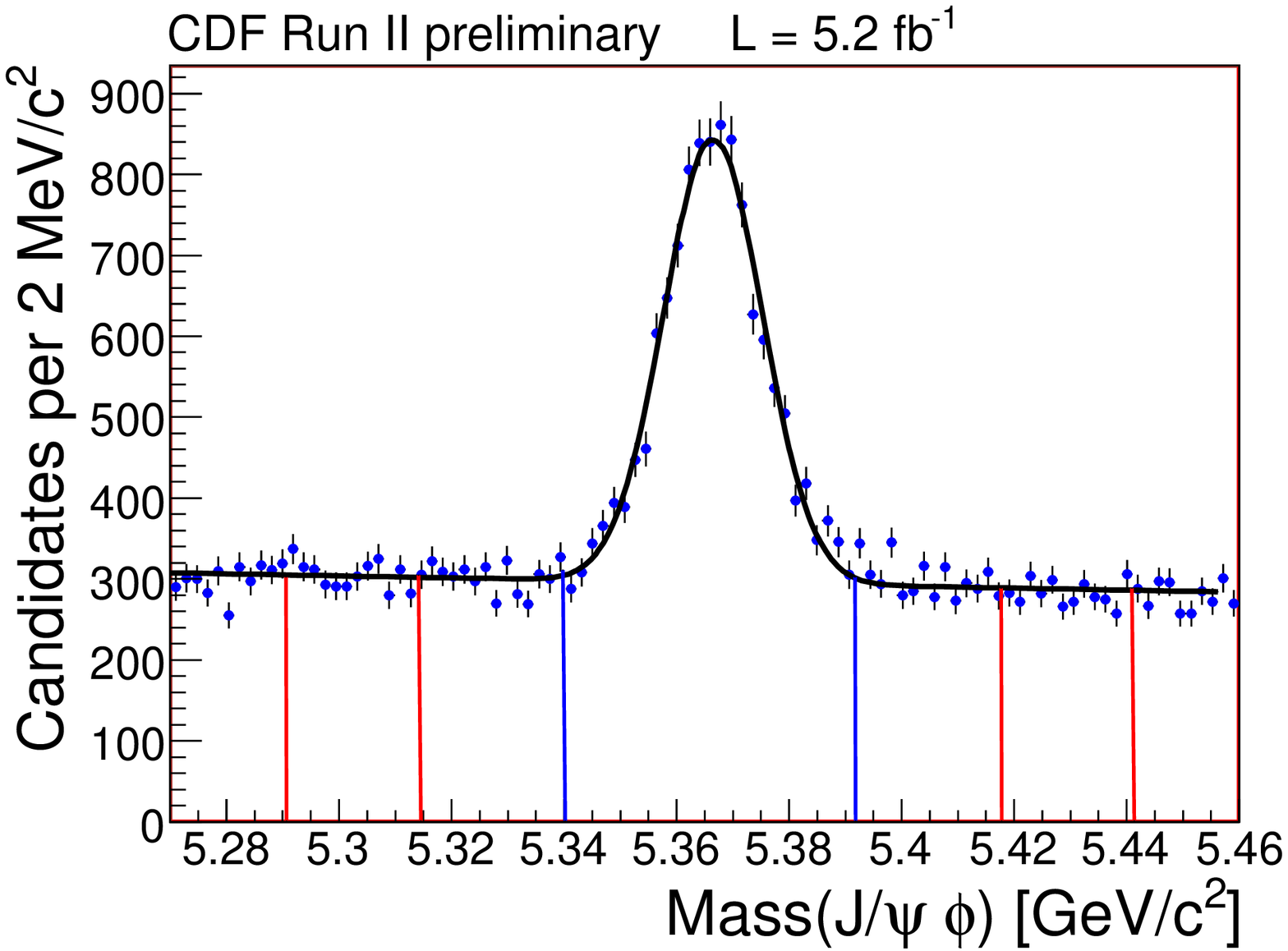}
\end{center}
%\vspace*{8pt}
\caption{Example of the dependence of the uncertainty on the \CP violating phase as a function of 
the requirement
on the neural network output at CDF (left). The invariant mass distribution of selected \BsJpsiphi
candidates at CDF (right).  \protect\label{fig1}}
\end{figure*}
An example of the outcome of simulation for the standard model \CP violation is shown in
Fig.~\ref{fig1}. The simulations which included different size of the \CP violation provide the same picture. 
It is interesting to note that the resulting selection accepts more background that the procedure used in 
early CDF analysis.
\begin{figure}[pht]
%\begin{minipage}{0.47\textwidth}
\includegraphics[height=3.0in,angle=-90]{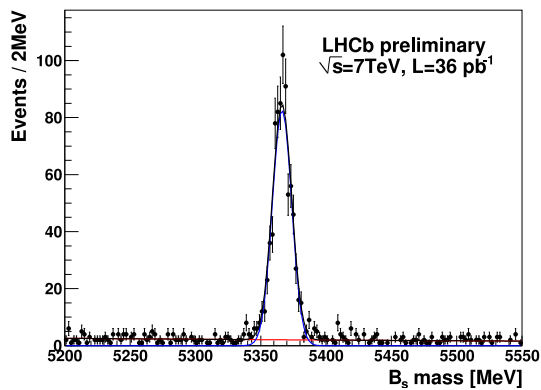}
%\vspace*{8pt}
\caption{The invariant mass distribution of the selected \BsJpsiphi 
candidates at LHCb. \protect\label{fig2}}
\end{figure}
%\end{minipage}
%\hfill
%\begin{minipage}{0.47\textwidth}
\begin{figure}[pht]
\includegraphics[width=3.0in]{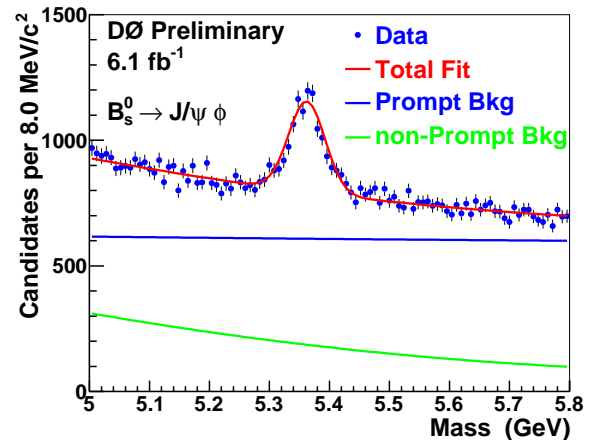}
%\vspace*{8pt}
\caption{The invariant mass distribution of the selected \BsJpsiphi 
candidates at D0. \protect\label{fig3}}
%\end{minipage}
\end{figure}
The selected sample comprises of $6504\pm85$ \BsJpsiphi signal events at CDF, $3435\pm84$ signal
events at D0 and $757\pm28$ \BsJpsiphi decays at LHCb. The invariant mass distributions from all
three experiments are shown in Figs.~\ref{fig1}--\ref{fig3}. It should be noted that LHCb selects
only candidates with proper decay time larger than 0.3 ps while the Tevatron experiments accept all
events independent of their decay time.

% Decay time reconstruction and typical resolutions (do I have D0 numbers?)
The decay time for each candidate is extracted from the displacement of the \Bsboth decay vertex away from
the primary vertex, which is reconstructed for each event separately. The uncertainty on the decay time
can be estimated separately for each candidate and this can be exploited in the analysis.
The typical resolution at the Tevatron
experiments is about 80--100 fs while thanks to the higher boost the resolution at LHCb is of the
order of 50 fs. The factor of 2 in the proper decay time resolution between Tevatron and LHCb gives
a significant advantage to the LHCb experiment in resolving the fast \Bsboth oscillations.

\section{Flavour tagging}
\label{sec:tagging}

As mentioned before, the important part of the analysis is to determine whether a given candidate was
produced as a \Bs or \Bsbar. This task is performed by algorithms called flavour tagging and is one of
the most challenging parts of the analysis.

% How flavour tagging works
At hadronic machines, most of the $b$ quarks are produced in pairs of opposite flavour (charge).
After hadronization, in the events interesting for the measurement, they end up in two independent
$b$-hadrons. Given that many other particles are produced together with the $b$-hadrons, one can
treat the two $b$-hadrons in the event as independent when considering their time evolution. This
independence splits possibilities to tag flavour into two basic classes. The first one, called same side
tagging, exploits the fragmentation process by which the \Bsboth meson is created out of the $b$ quark. The 
second class, called opposite side tagging, uses properties of the decay of the $b$-hadron which contains
the other $b$ quark. To fully characterize the performance of a flavour tagging algorithm, two quantities
are needed. The first one is the efficiency, $\epsilon$, which gives the fraction of the candidates for which
a flavour tagging decision is made. The second one, called dilution, $D$, provides the information 
about the chance that the decision made is correct. Formally it is defined as
\begin{equation}
D=\frac{N_{\mathrm{R}}-N_{\mathrm{W}}}{N_{\mathrm{R}}+N_{\mathrm{W}}}=2P_{\mathrm{R}}-1,
\end{equation} 
where $N_{\mathrm{R}}$ ($N_{\mathrm{W}}$) is the number of correct (wrong) decisions and
$P_{\mathrm{R}}$ is the probability of having the correct decision. Often the two performance quantities
are combined into a single value as $\epsilon D^2$, which gives the effective statistics of the tagged
sample. If we have $N$ events with given flavour tagging performance, they will be equivalent to
$N\epsilon D^2$ events with perfectly known production flavour. Typically the dilution is estimated on
a candidate-by-candidate basis, and often used as additional information in the fits in order to
increase the sensitivity to the oscillation behaviour of the \Bsboth.

\subsection{Flavour tagging algorithms}

The physics idea behind the same side tagging is rather simple. In order to produce a \Bs, one needs to
attach a strange quark to the bottom anti-quark in the hadronization process. The strange quark
normally originates from a pair of quark and anti-quark and thus the remaining strange anti-quark has to
end up in another hadron. With significant chance this other hadron is a kaon. If it is a charged kaon
and the experimentalist succeeds in picking-up the correct corresponding track, then the charge of the track
determines also the flavour of the \Bs at the production time. While the general idea is rather simple, the
details of the hadronization are far from being fully understood and therefore it is hard to develop
such an algorithm. The selection of the tagging track can be done based on kinematical and/or particle
identification information. From a point of view of kinematics, the track which carries the 
fragmentation partner will typically have small transverse momentum relative to the \Bsboth and a large 
component of momentum in the direction of the \Bsboth. The particle identification is a significant 
help as most of the particles produced in $\Pproton\Pproton$ or $\Pp\Pap$ collisions are pions and therefore 
a track positively identified as a kaon and being close to the \Bsboth has a large probability to be the 
right track.

The opposite side algorithms exploit the decay of the second $b$-hadron in the event. Once its flavour
is determined, as the bottom quarks are produced in pairs of quark and anti-quark, the production
flavour of the detected \Bsboth can be taken as opposite.

The most common way of identifying the flavour of the
other $b$-hadron is to exploit semileptonic decays. They have a rather large branching fraction and
provide a unique and clear experimental signature. The charge of the lepton is directly
correlated to the flavour of the $b$ quark in the decaying hadron. All experiments use decays which contain
either an electron or muon. As the leptons are dominantly produced by the decays of heavy flavour ($b$ and 
$c$) quarks, there is a large chance that the identified lepton originates in the $b$-hadron decay.

The second piece of information which can be used for opposite-side tagging is based on the $b\rightarrow 
c\rightarrow s$ decay chain of the $b$
quark with the strange quark ending up in kaon. If charged, it contains information about the flavour
of the decaying $b$-hadron. Experimentally one is looking for a charged kaon, which does not point
to the primary vertex. 

The other piece of information to use is more inclusive and utilizes a jet charge by CDF
experiment and a secondary vertex charge by D0 and LHCb. The jet charge is calculated for each jet as
\begin{equation}
Q_\mathrm{jet}=\sum q_ip_T^iw_\mathrm{NN}^i/\sum p_T^iw_\mathrm{NN}^i,
\end{equation}
where sums run over all tracks associated with the jet and $q_i$ is
the charge of the $i$-th track, $p_T^i$ is the transverse momentum of the given track, and $w_{NN}^{i}$ is
the probability that the track originates from the $b$-hadron decay \cite{Lecci:2005gf}.
The calculated jet charge is then used to decide on the production flavour of the \Bsboth meson.
Three classes of jets are distinguished at CDF. The first class contains events with a secondary vertex
within the jet. The second class is defined to contain events without a secondary vertex, but at least one
track with a significant probability to come from $b$-hadron decay and the third class contains all
events failing the criteria for the first two classes.  The reason for the distinction is that different 
classes have different performance and distinguishing them helps to optimize the overall performance.
The secondary vertex charge uses a similar strategy, but instead of using the full jet, it exploits tracks
assigned to the secondary vertex. The charge is calculated as
\begin{equation}
Q_{SV}=\frac{\sum q_iw_i}{\sum w_i},
\end{equation}
where $q_i$ is the charge of the track and $w_i$ is the weight. There are several options for weights which
are in principle equivalent. The D0 experiment uses as $w_i$, the longitudinal momentum of the track
along the direction of total momentum of the tracks assigned to the secondary vertex
\cite{Abazov:2006qp}.
LHCb on the other hand uses as weight the transverse momentum of the track to the power of
0.4 \cite{LHCb-2007-058}.

In each of the three experiments the different algorithms are executed separately and their outputs
are then combined together. Typically in the combinations, opposite side algorithms are combined to
a single decision and then the same side algorithm, if used, is handled as uncorrelated to the opposite
side decision. For combination of opposite side algorithms, CDF uses a neural network while D0 and
LHCb use a likelihood method.

%\subsection{Calibration and performance}
\subsection{Calibration}
% How it can be calibrated
All three experiments make an effort to calibrate the mistag probabilities directly on the data. It is
useful to treat the opposite- and the same-side tagging separately. As the opposite side tagging is
independent of the reconstructed $b$-hadron, it is possible to use the more abundant \PBpm and
\PBzero mesons for the calibration. On the other hand, same-side tagging depends on the $b$-meson under study
and therefore has to be calibrated using the \Bsboth mesons for the application discussed here.

\subsubsection{Opposite side tagging}
\begin{figure*}[pht]
\includegraphics[width=3.0in]{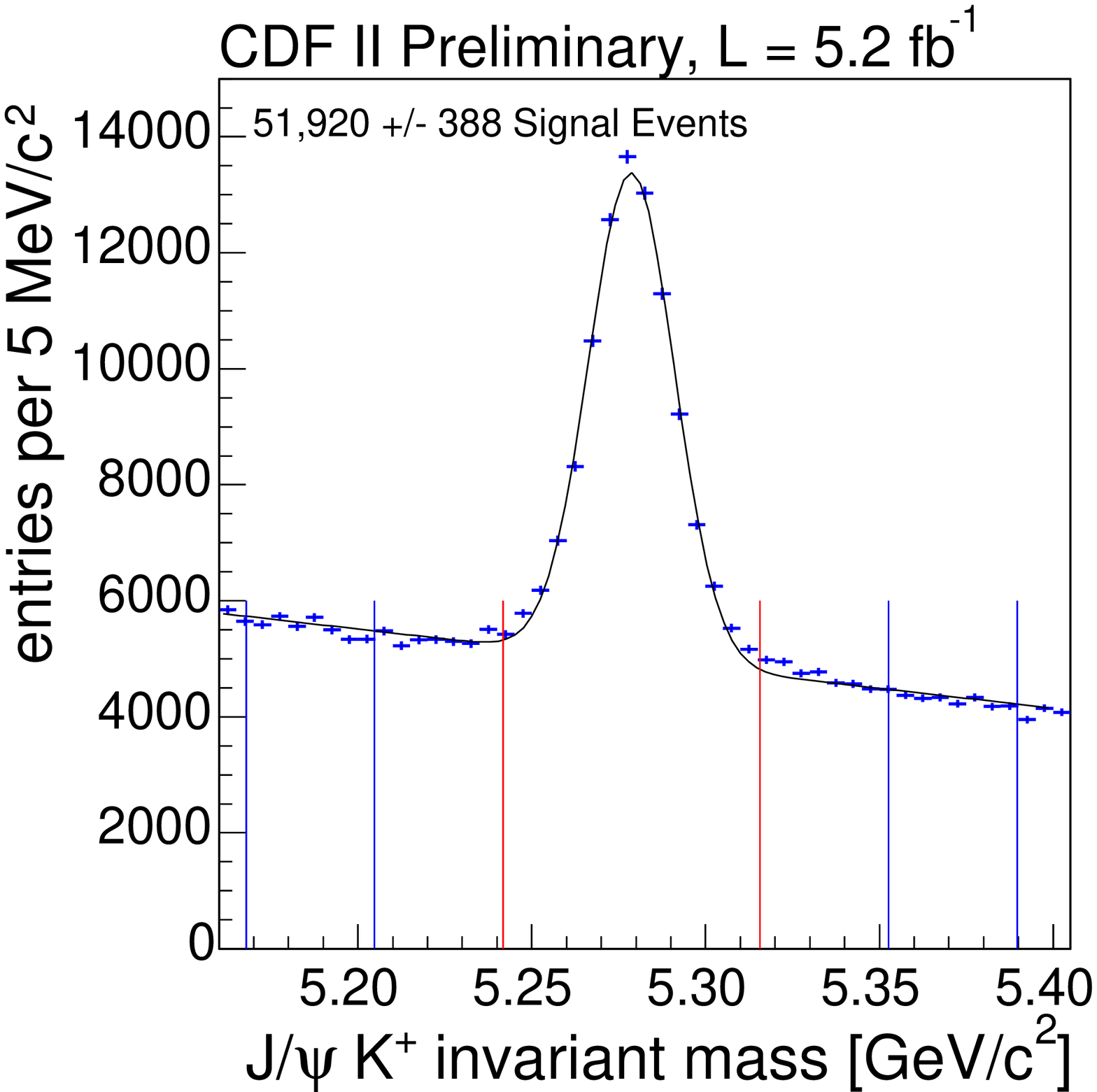}
\includegraphics[width=3.0in]{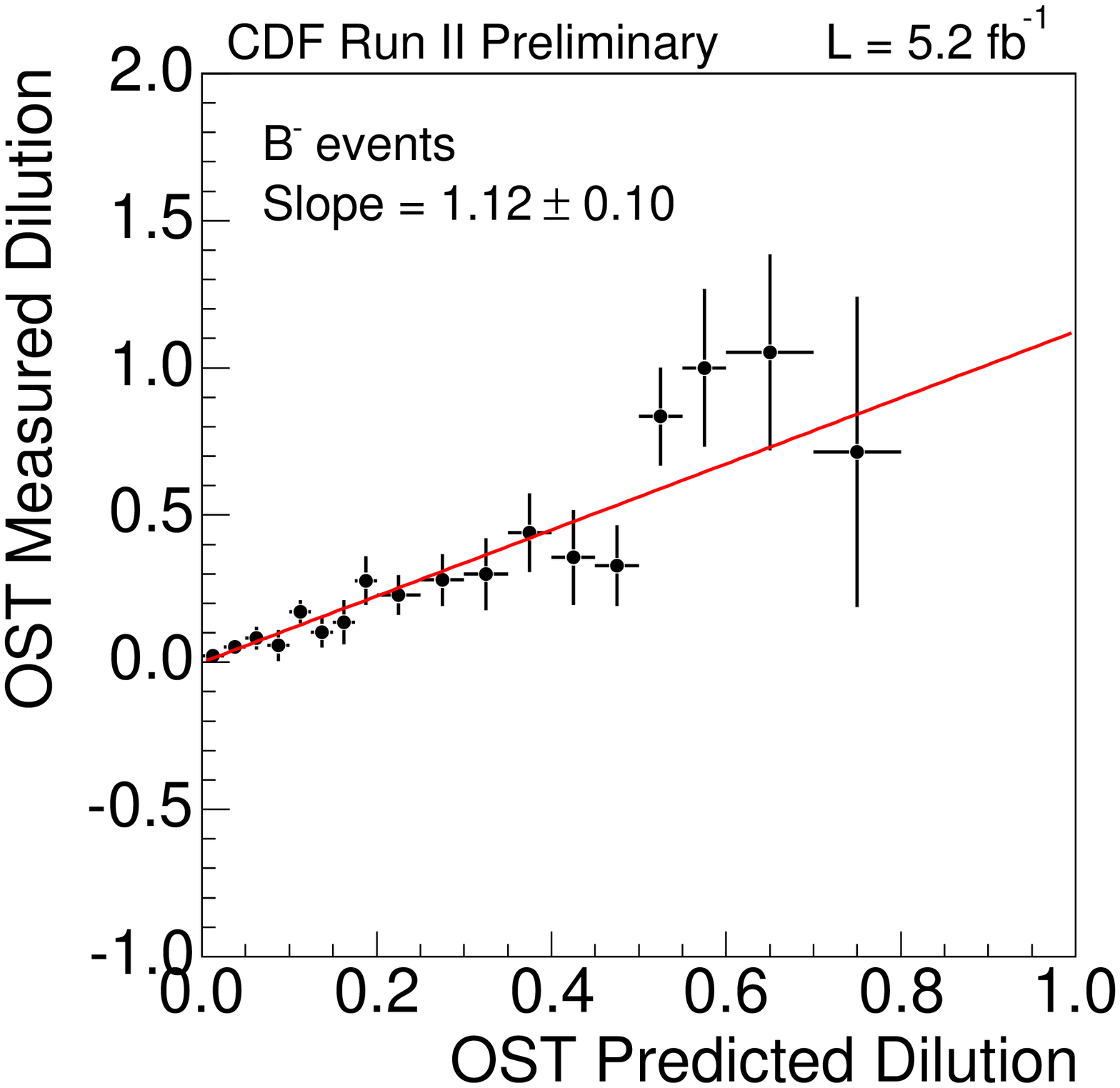}
%\centerline{\psfig{file=figs/Bp_mass.eps,width=2.0in}
%            \psfig{file=figs/OST_Bminus_all.eps,width=2.0in}}
%\vspace*{8pt}
\caption{Example of the opposite side flavour tagging calibration using $\PBpm\rightarrow \Jpsi \PKpm$
decays at CDF. On the left we show the $\Jpsi \PKpm$ invariant mass distribution showing clear \PBpm signal.
On the right we show the measured dilution as a function of the estimated dilution for the \PBm
only.  \protect\label{fig:OSTcalibration}}
\end{figure*}

The easiest way of calibrating the opposite side tagging algorithms is to use fully reconstructed
\PBpm decays. In the context of the measurement of the \CP violation in \BsJpsiphi decays, the most useful decay is
$\PBpm\rightarrow \Jpsi \PKpm$. The advantage of the \PBpm is given by the fact that its reconstruction
efficiency is rather high due to the lower number of tracks needed in the reconstruction and more
importantly that it does not oscillate. Therefore the charge obtained from the decay products unambiguously
identifies the flavour at the production time. As the production flavour is known for each
candidate, one can easily measure the probability $P_\mathrm{R}$ of having a correct decision and compare 
this with the probability estimated by the algorithm. The high abundance of the reconstructed \PBpm
signal also allows the measurement of the flavour tagging asymmetry by splitting the sample into \PBp and 
\PBm. As an example, Fig.~\ref{fig:OSTcalibration} shows the invariant mass distribution of reconstructed
\PBpm signal at CDF and the dependence of the measured dilution on the estimated dilution.
In this case, the ideal behaviour is a linear dependence with a slope of unity. The slope itself is later 
used in the fit for the \CP violation in \BsJpsiphi to correct the estimated dilutions of the opposite-side 
taggers.

 The second option for the calibration of the opposite-side tagging is to use \PBzero decays and 
measure the oscillation pattern. It consists of measurement of the asymmetry
\begin{equation}
A_{\mathrm{mixing}}(t)=\frac{N_{\mathrm{mix}}(t)-N_{\mathrm{unmix}}(t)}{N_{\mathrm{mix}}(t)+N_{\mathrm{unmix}}(t)}
\end{equation}
where $N_{\mathrm{mix}}$ ($N_{\mathrm{unmix}}$) is the number of \PBzero candidates with the same (opposite)
flavour at the production and decay time. The amplitude of the asymmetry directly encodes the
performance of the flavour tagging algorithm. While the achievable precision cannot compete with
the \PBpm decays, it provides a demonstration that particle oscillations can be resolved and thus builds
up the overall confidence in the analysis. While each of the three experiments performed the \PBzero mixing
measurement at some point on the way to current work on \CP violation in \Bs, it is not widely used
in the latest results due to its statistical limitations.

\subsubsection{Same side tagging}

A more difficult part is to calibrate the same side flavour tagging as it can be done only by using
the \Bsboth itself. The principle is the same as using \PBzero to calibrate the opposite side tagging. 
Measuring the \Bsboth oscillations using flavour-specific \Bsboth decays, the asymmetry amplitude provides 
direct information on the flavour tagging power. The difficulty is in the lower \Bsboth yield and the very 
fast oscillations, which make it hard
to obtain a significant mixing signal. In addition, care has to be taken to properly describe the decay time
resolution as its mis-modelling also affects the mixing asymmetry amplitude. On the other hand, one could
argue that even if there is some decay time resolution mis-modeling, it is likely to be the same as
in the case of decay \BsJpsiphi and therefore does not pose a real issue for the analysis. 

To calibrate the same-side tagging, CDF uses
the decays $\Bsboth\rightarrow \PDsmp\Ppipm$ and $\Bsboth\rightarrow\PDsmp\Ppipm\Ppiplus\Ppiminus$. The
\PDsminus is
reconstructed in decays to $\Pphi\Ppiminus$, $\PKst\PKminus$ or $\Ppiplus\Ppiminus\Ppiminus$, where for
the decay $\Bsboth\rightarrow\PDsmp\Ppipm\Ppiplus\Ppiminus$, only the $\PDsminus\rightarrow\Pphi\Ppipm$ is
used. Altogether about
12900 signal events are reconstructed. The same-side flavour tagging algorithm estimates for each
candidate dilution, which is taken into account in the \Bsboth oscillation fit. From the measured amplitude of
the \Bsboth mixing, CDF derives a single scaling factor for the estimated dilution of $A=0.94 \pm 0.15
(\mathrm{stat}) \pm 0.13 (\mathrm{syst})$ \cite{Aaltonen:2010ss}. In this approach, one cannot correct the 
shape of the dilution distribution, but only adjust the overall average scale of it. 

% LHCb SSKT calibration
The LHCb experiment plans to use the \Bsboth mixing to calibrate the same side tagging. The first study using
data collected in 2010 allowed the measurement of the \Bsboth mixing frequency using the opposite-side 
flavour tagging \cite{LHCb:BsMixing}. While the measurement of the \Bs mixing was possible, the same side 
flavour
tagging performance is not sufficient to obtain a significant signal with the same-side flavour tagging
only. As a consequence the first measurement of the \CP violation in \BsJpsiphi from the LHCb experiment
does not use the same-side flavour tagging. 

% D0 SST calibration note
The situation at D0 is different as there is no effective way to trigger on the hadronic
decays. Therefore it is impossible to obtain sufficient statistics in the fully reconstructed
flavour-specific decays.
There is the possibility of reconstructing a large sample of semileptonic decays but due to the missing
neutrino, the time resolution is significantly worse, which makes it difficult to obtain a significant
\Bsboth mixing signal. All this makes calibration of the same-side tagging at D0 very difficult and the D0
experiment did not attempt to perform it up to now. It should be noted that while D0 used the same-side
tagging in previous rounds of the analysis, the latest analysis does not use same-side flavour
tagging.

% Performance
\subsection{Performance}

The performance of the flavour tagging at the CDF experiment is $\epsilon D^2\approx 1.2\%$ for the 
opposite-side algorithm and $\epsilon D^2\approx 3.2\%$ for the same-side algorithm. As there is an overlap
between the two taggers, it is not straightforward to combine the two numbers into the single
performance number.  The D0 experiment achieves
for its opposite-side flavour tagging $\epsilon D^2\approx 2.5\%$. Finally the opposite-side flavour
tagging performance at the LHCb experiment is $\epsilon D^2\approx 2.2\%$. 

It is interesting to compare the CDF
and D0 performances as the two detectors cover basically the same phase space region and work in the same
environment. The factor two in the performance of the opposite side flavour tagging comes almost
entirely from the better muon system, which has larger coverage at D0 compared to CDF, but also lower
misidentification rate. On the other hand, CDF benefits from limited particle identification, which 
significantly boosts the performance of the same-side flavour tagging.  The main drawback of the same-side
flavour tagging is in calibration, which requires significant effort for quite limited
precision. 

The comparison between the Tevatron and LHCb experiments is more difficult as they cover
different phase space regions. Despite that we can still safely say that LHCb strongly benefits from
the excellent particle identification, which boosts their opposite-side flavour tagging by identification of
kaons coming from the $b\rightarrow c\rightarrow s$ decay chain. Also, thanks to the forward geometry and
design specific for flavour physics, the identification of electrons should be easier compared to
the multi-purpose detectors at Tevatron. At the same time the same-side flavour tagging at LHCb
is expected to perform worse than at CDF as the track density in the forward region is larger and thus it is
more difficult to pick up the correct track. It should be also noted that in 2010, the LHCb experiment was
taking data with the number of interactions per bunch crossing well above the design value, which means again
a more difficult environment for the flavour tagging.

\section{Fit description}
\label{sec:fit}

It is time to get to the heart of the analysis, which is a maximum likelihood fit. We are going to
skip most of the details of the background description and refer the reader to the original work of the
experiments. In short, the background is described using a phenomenological description derived mostly
using data events in the \Bsboth mass sidebands. The important part we want to discuss in detail is
the description of the \Bsboth decays in the fit. It is rather instructive to read through the details in 
Refs.~\cite{Azfar:2010nz,Xie:2009fs}. The CDF analysis is based fully on the description in 
Ref.~\cite{Azfar:2010nz}. As we will discuss in the following, the LHCb and D0 analyses do not implement 
all the subtleties of the decay.

As we have already discussed, for each candidate which the experiments reconstruct, the flavour tagging 
determines whether the candidate was produced as a \Bs or \Bsbar. In the following, this is encoded in the 
variable $\xi$, which takes values 1 for the \Bs, -1 for the \Bsbar, and 0 if the flavour tagging is unable 
to make a decision. The signal probability function is given by the weighted average of the probability 
density functions for \Bs and \Bsbar which takes the form
\begin{eqnarray}
  P_s(t,\vec\rho,\xi|{ D},\sigma_{t}) 
      &=& \frac{1+\xi{ D}}{2} P(t,\vec\rho| \sigma_{t}) \epsilon(\vec\rho)\nonumber \\
      &+&\frac{1-\xi{ D}}{2}\overline{P}(t, \vec\rho| \sigma_{t}) \epsilon(\vec\rho), 
\end{eqnarray}
where $P$ and $\overline{P}$ are the probability density functions for \Bs and \Bsbar. The quantities $t$ and
$\sigma_t$ are the decay time and its uncertainty for a given candidate, $\vec\rho$ contains the measured
decay angles in the transversity basis \cite{Dighe:1995pd} and $D$ is the dilution predicted for the given 
candidate. Finally $\epsilon(\vec\rho)$ parametrizes the angular efficiency. It should be noted that each of 
the experiments has some minimum requirements on the momentum in order to be able to reconstruct a track
and this requirement reflects in non-uniform angular efficiencies. This minimal requirement is a consequence 
of the geometry of the detector and is in principle very hard to avoid.

% P-wave part, what can be learned
The decay time and angular distribution of the decay \BsJpsiphi is given in
Refs.~\cite{Dighe:1995pd} as 
\begin{eqnarray}
\frac{\dd^4P(t,\vec\rho)}{\dd t\dd\vec\rho} 
&\propto& |A_0|^2 {{\cal T}_+} { f_1(\vec\rho)} + |A_\parallel|^2 {{\cal T}_+}
{ f_2(\vec\rho)} \nonumber \\ &+& |A_\perp|^2 {{\cal T}_-} { f_3(\vec\rho)} 
+ |A_\parallel||A_\perp| {{\cal U}_+} { f_4(\vec\rho)} \nonumber \\ &+& |A_0||A_\parallel|
\cos(\delta_{\parallel}) {{\cal T}_+} { f_5(\vec\rho)}\nonumber \\
&+& |A_0||A_\perp| {{\cal V}_+} { f_6(\vec\rho)},
\end{eqnarray}
where $|A_0|$, $|A_{||}|$ and $|A_\perp|$ are three polarization amplitudes and the functions
$f_i(\vec\rho)$ describe the angular distributions \cite{Dighe:1995pd}. The first three terms describe 
amplitudes-squared while the other three terms describe interferences between the three amplitudes. 
The description of \Bsbar is obtained by substituting ${\cal U}_+$ by 
${\cal U}_-$ and ${\cal V}_+$ by ${\cal V}_-$.
The functions ${\cal T}_\pm$, ${\cal U}_\pm$ and ${\cal V}_\pm$ provide the time dependence and take the form
\begin{eqnarray}
{\cal T}_{\pm}=e^{-\Gamma t}&\times&\left[\cosh(\Delta \Gamma t/2)
                 \mp { \cos(\phiS)\sinh(\Delta
\Gamma t/2)}\right.\nonumber\\ & &\left.{ \pm\ { \eta\sin(\phiS)\sin(\Delta m_st)}}\right], \\
{\cal U}_{\pm} =\pm e^{-\Gamma t}&\times&\left[{\sin(\delta_{\perp} -\delta_{\parallel})\cos(\Delta
m_st)} \right.\nonumber \\  &-&\left.{ \cos(\delta_{\perp}-\delta_{\parallel})\cos(\phiS)\sin(\Delta m_st)}\right. \nonumber \\
               &\mp&\left.{
\cos(\delta_{\perp}-\delta_{\parallel})\sin(\phiS)\sinh(\Delta\Gamma t/2)}
                  \right], \\
{\cal V}_{\pm} =\pm e^{-\Gamma t}&\times&\left[{\sin(\delta_{\perp})\cos(\Delta m_st)}\right. \nonumber\\
               &-&\left.{\cos(\delta_{\perp})\cos(\phiS)\sin(\Delta m_st)}\right.\nonumber  \\
               &\mp&\left.{\cos(\delta_{\perp})\sin(\phiS)\sinh(\Delta\Gamma t/2)}\right].
\end{eqnarray}
In this, $\delta_{||}$ and $\delta_\perp$ are strong phases between amplitudes and $\eta$ is 1 for
\Bs and $-1$ for \Bsbar.

There are a few interesting points to note which are different in this case compared to the analogous 
analysis of the \CP violation in $B^0\rightarrow\Jpsi K_s^0$. First if we consider the case without flavour 
tagging, which corresponds to the case of $\xi=0$ for each event, the terms $\sin(\Delta m_s t)$ cancel out, 
but there are several other terms which are sensitive to $\phiS$. This comes from the fact that both
\Jpsi and $\phi$ are spin 1 particles and thus we deal with a mixture of \CP-even and \CP-odd final
states, which interferes and the non-zero $\Delta\Gamma$. As a benefit from this complexity,
the system provides an additional sensitivity to the \CP violation from the interference terms, which is available
even without flavour tagging. The second point to note is that even if we put the \CP violation to zero,
equivalent to $\sin(\phiS)=0$, the interference between \CP-even and \CP-odd provides sensitivity to the
\Bsboth mixing frequency \cite{Azfar:2010nz} which can be exploited by experiments like ATLAS and CMS to perform
a measurement also without access to the hadronic \Bsboth decays. Finally, in the case of no \CP
violation one is sensitive to the strong phase $\delta_\perp$ only with the flavour tagging, while
without the flavour tagging, $\delta_\perp$ is unaccessible. 

\begin{figure}[pht]
%\begin{minipage}{0.47\textwidth}
\includegraphics[width=3.0in]{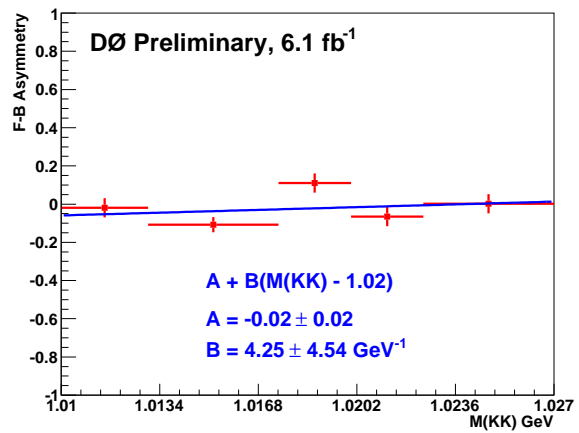}
%\centerline{\psfig{file=figs/B60F14,width=2.0in}}
%\vspace*{8pt}
\caption{The forward-backward asymmetry of the positive kaon angular distribution across the invariant mass
of the kaon pair. In the absence of the s-wave, one expects a flat dependence, while in the presence, the 
dependence becomes non-trivial.
 \protect\label{fig:D0AFB}}
\end{figure}

% Issue of s-wave
A delicate issue is the question of a possible s-wave contribution to the reconstructed \Bsboth signal. While 
each experiment has rather tight selection on the invariant mass of the kaon pair around the world average 
$\phi$ mass, decays like nonresonant $\Bsboth\rightarrow \Jpsi \PKplus\PKminus$ or \BsJpsifnot with 
$\Pfz\rightarrow \PKplus\PKminus$ can contribute as well. Original estimates of the branching fraction 
relative to \BsJpsiphi decay 
\begin{equation}
R_{f_0/\phi}=\frac{\mathcal{B}(\BsJpsifnot)}{\mathcal{B}(\BsJpsiphi)}
\frac{\mathcal{B}(\Pfz\rightarrow
\Ppiplus\Ppiminus)}{\mathcal{B}(\Pphi\rightarrow
\PKplus\PKminus)}
\end{equation}
yielded values around 0.2 \cite{Stone:2008ak}. While this estimate did not include the effect of the selection 
on the kaon pair invariant mass, it was argued that neglecting the s-wave contribution can bias the result 
for the \CP violation \cite{Stone:2010dp}. Recent observation of the decay \BsJpsifnot and measurement of 
$R_{{f_0/\phi}}$ to be about 0.25 \cite{Aaij:2011fx,Li:2011pg,Aaltonen:2011nk,Abazov:2011hg} 
further support the necessity to take the s-wave contribution into 
account in some way. With a typical selection on the invariant mass of kaon pairs this would
translate to about a 1.5\% contribution from the \BsJpsifnot decay in the selected sample with rather
large uncertainties due to badly known \Pfz branching fractions. 
At this stage the most complete treatment is done by the CDF experiment and uses
the formalism from Ref.~\cite{Azfar:2010nz}. The analysis incorporates an additional amplitude yielding
four more angular terms, one for the s-wave amplitude squared and three for the interference between the 
original amplitudes of the \BsJpsiphi decay and the s-wave amplitude. The implementation treats the invariant
mass of the kaon pair as an unobserved variable. While the s-wave complicates an already complex analysis
further, it might provide an additional benefit in helping to resolve ambiguities in the value of the 
\CP-violating phase. In contrast to the CDF treatment, neither D0 nor LHCb implements at this stage the 
s-wave contribution into the fit. As the presence of the significant s-wave contribution would introduce some
asymmetry in the distribution of the kaon angle, experimentally it is possible to check for its
presence. This was for the first time seen in the \BzeroJpsiKS decays at the Babar
experiment \cite{Aubert:2007hz}. The
D0 experiment performs such a check by inspecting the forward-backward asymmetry in the kaon angle
distribution in five different intervals of the kaon pair invariant mass (see Fig.~\ref{fig:D0AFB}). 
%\end{minipage}
%\hfill
%\begin{minipage}{0.47\textwidth}
\begin{figure}[pht]
\includegraphics[width=3.0in]{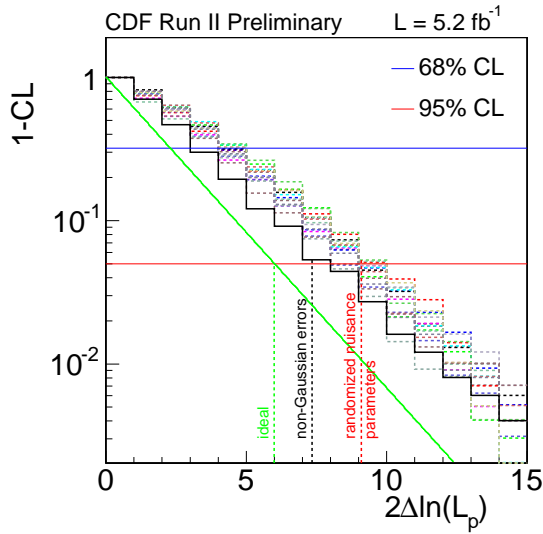}
%\centerline{\includegraphics[width=2.0in]{figs/tag_2d_coverage}}
%\vspace*{8pt}
\caption{The map between actual coverage and the likelihood ratio for the CDF measurement. The continuous
line shows ideal Gaussian behaviour, the full histogram the default set of generated experiments and
the dashed histograms show 16 alternative sets.
 \protect\label{fig6}}
%\end{minipage}
\end{figure}%
From this they conclude that no significant s-wave is present and neglect it in the rest of the analysis. 
On the other hand, LHCb neglects s-wave contribution in the fit, but evaluates the systematic uncertainty 
using information on the s-wave derived by the CDF experiment in their full fit. Thus while none of the
experiments has evidence for non-zero s-wave contribution, the CDF and LHCb experiments include such a 
possibility to the uncertainties, while D0 remains more aggressive and does not assign any
uncertainties for the possible s-wave contribution.

\section{Statistical issues}
\label{sec:statistics}

% Statistical issues in the analysis
While the interference between \CP-odd and \CP-even amplitudes helps in the determination of the \CP
violating phase \phiS, it also introduces some non-trivial statistical issues.
As is apparent from Sec.~\ref{sec:fit} that the probability density function
contains several periodic functions, resulting in some symmetries if we consider the \BsJpsiphi
decays without the s-wave contribution. Neglecting the s-wave contribution, the system is invariant under
simultaneous transformation $\phiS\rightarrow\pi-\phiS$, $\Delta\Gamma\rightarrow
-\Delta\Gamma$, $\delta_\perp\rightarrow \pi-\delta_\perp$, and $\delta_{||}\rightarrow
2\pi-\delta_{||}$. As a consequence there are two equivalent solutions which in the case of small
statistics are not well separated. This fact itself makes minimization of the likelihood a difficult
task. The issue of symmetries can appear not only for \phiS, but also for strong phases and, if their true
values are close to the symmetry point, their extraction is again non-trivial.

As was discussed before, there is a possibility to extract information on the \CP violation even without
flavour tagging. In this case, the situation becomes even more difficult. The single symmetry of the
flavour tagged case turns to two independent symmetries, which are $(\phiS\rightarrow
-\phiS,\,\delta_\perp\rightarrow\delta_\perp+\pi)$ and $(\Delta\Gamma\rightarrow -\Delta\Gamma,\,
\phiS\rightarrow \phiS-\pi)$. The consequence is the existence of four solutions compared
to two in the flavour tagged case. An additional complication arises from the fact that the strong phase
$\delta_\perp$ appears always in a product with $\sin(\phiS)$. As a result, in case of no \CP
violation there is no sensitivity to $\delta_\perp$, but if the sensitivity to \CP violation is small,
the fit tends to bias the result as by increasing the \CP violation, the fit gains $\delta_\perp$ as an 
additional parameter available to describe the statistical fluctuations. Moreover the bias is non-linear and
decreases with increasing true \CP violation.

% How frequentists result is obtained (add differences between experiments)
It follows from the symmetries that there is a danger of non-Gaussian behaviour of the likelihood,
which to some extent depends on the statistics and the true values of the parameters. If the true values 
are close to the symmetry points, more statistics are needed to clearly resolve those. Given the
importance of the measurement for putting bounds on new physics, it is important to make sure that
any non-Gaussian behaviour is properly taken into account. In order to achieve this, the experiments
resort to a frequentist treatment based on the likelihood ratio ordering suggested by Feldman and
Cousins \cite{Feldman:1997qc}. To construct confidence level regions in the $\phiS$--$\Delta\Gamma$ plane, 
the procedure is to evaluate for each point, the ratio of likelihoods between the fit with \phiS and 
$\Delta\Gamma$ fixed to specific values, and the fit where they are allowed to float. The likelihood 
ratio is then compared to the set of simulated experiments. For each point a p-value is obtained as a 
fraction of the number of simulated experiments which have a likelihood ratio larger than the one observed in 
the data. Connecting points with the same p-value yields the corresponding confidence-level region.

The difficulty in the procedure is that there are many parameters involved, for which we do not know the true
values, but only estimates from experiments which could be from the analysis itself or some
measurement external to the analysis. To deal with this, CDF generates one set of experiments using
the values of all parameters except \phiS and $\Delta\Gamma$ from the global minimum of the
likelihood together with 16 alternative sets where all parameters are chosen randomly from
a $\pm5\sigma$ hypercube around the global minimum. An example of the map between the actual coverage and
the likelihood ratio for the CDF experiment is shown in Fig.~\ref{fig6}. 
While the procedure does not guarantee the exact coverage of the derived contours, it is assured that there 
is no undercoverage. The LHCb experiment employs a similar procedure, but from the available information it 
is not fully clear to what extent they vary the input parameters in the simulated experiments. The D0 
experiment decides to take a different path and rather than going through a full frequentist treatment, 
they constrain the strong phases to the values
measured in the \BzeroJpsiKS decay. This point is rather controversial with theoretical arguments
supporting it presented in Ref.~\cite{Gronau:2008hb}, but the argument is generally not fully accepted. From
the experimental point of view, the constraint effectively removes some symmetries and better separates two
minima. This results in a likelihood which is closer to the Gaussian shape, but still needs a small
adjustment, which is performed in a similar way to CDF and LHCb.

To finish the discussion on the statistical issues, a note on the importance of the flavour tagging and the 
time resolution is in order. As we discussed, there are two pieces of information about the \CP violation in 
the analysis. The first one is in the interference between \CP-even and \CP-odd amplitudes and this one does 
not require the resolution of the \Bsboth oscillations. Therefore the time resolution for this part is not 
critical. On the other hand, the importance is driven by the size of $\Delta\Gamma$. With larger 
$\Delta\Gamma$, the importance increases. As a consequence, there is some correlation between the uncertainty 
on \phiS and the value of $\Delta\Gamma$ extracted by a given experiment. The second part of the sensitivity
comes from resolving the \Bsboth oscillations and here the flavour tagging performance and the time 
resolution are crucial. Here LHCb has the clear benefit of better time resolution compared to the Tevatron
experiments. On the other hand, the flavour tagging performance is behind the CDF experiment at this
stage and smears out part of the benefit from their better decay time resolution.
% Do we need anything else?
% Things not to forget in this section:
%   - Dependence on sensitivity from interference on DG
%   - Importance of flavour tagging on t resolution and DG

\section{Results}
\label{sec:results}

% Provide short list of results on CPV phase -- Delta Gamma
Typically in each case two different fits are performed. One in which no \CP violation is assumed
and the values of physics parameters such as mean lifetime, decay width difference, and amplitudes are
measured, and the second fit in which constraints on \CP violation are derived. The main result of the 
analysis is given as bounds in the $\Delta\Gamma$--$\phiS$ plane.

\begin{figure*}[pht]
\includegraphics[width=3.0in]{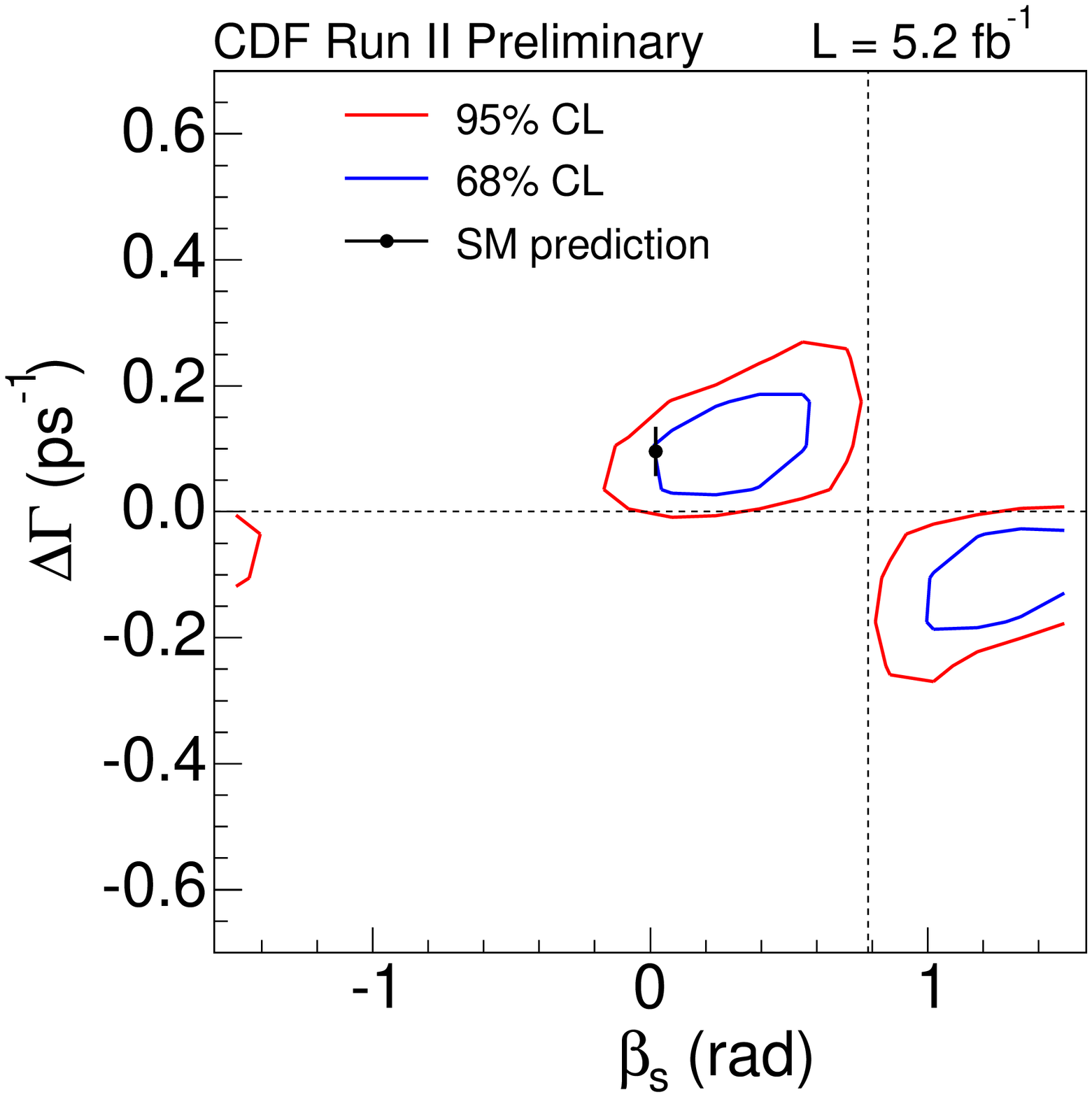}
\includegraphics[width=3.0in]{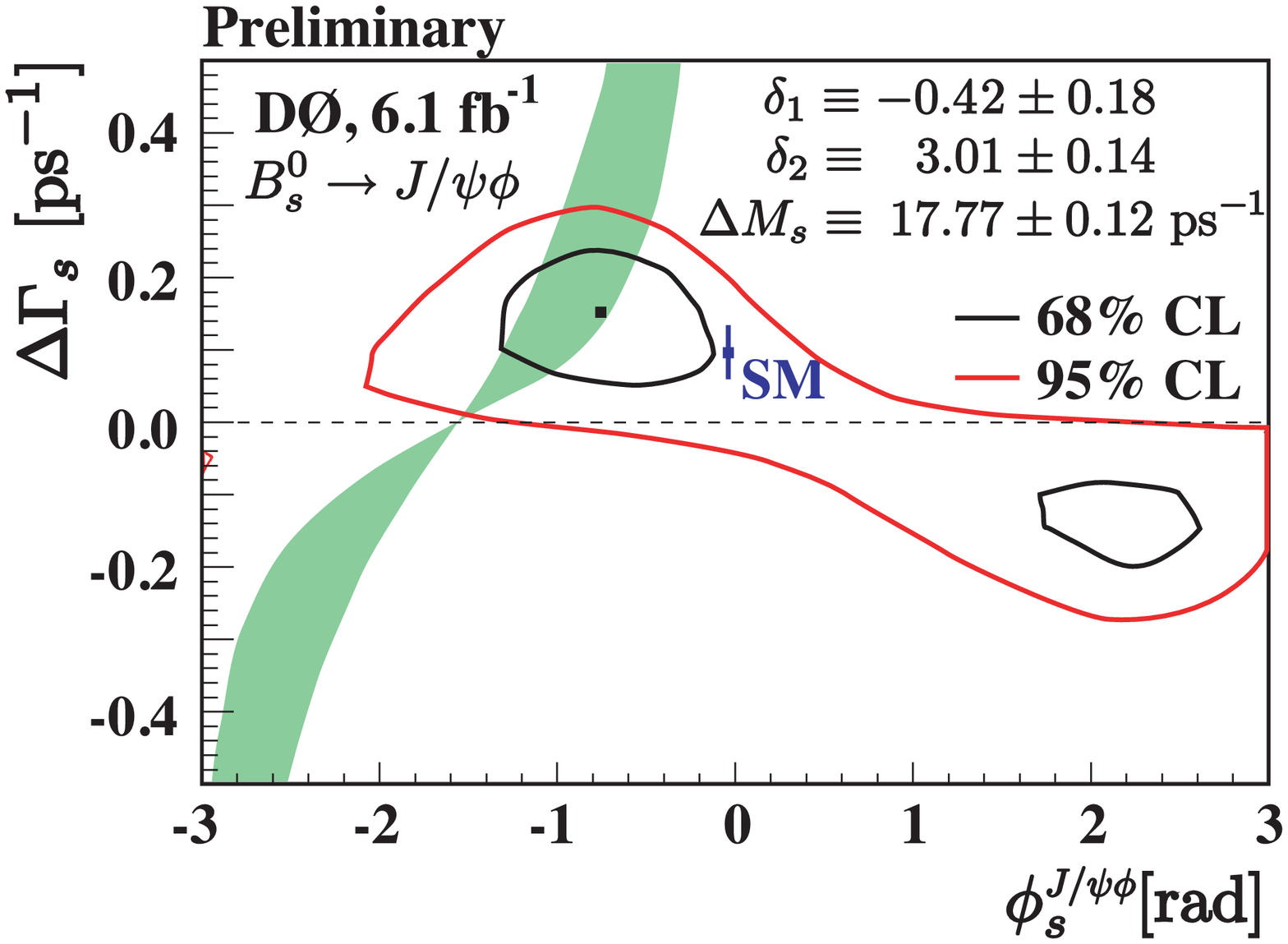}
%\centerline{\psfig{file=figs/tag_contour_systadjust,width=2.0in}\hspace{1cm}
%            \psfig{file=figs/B60F10,width=2.0in}}
%\vspace*{8pt}
\caption{The bounds in the $\Delta\Gamma$--$\phiS$ plane obtained by the CDF (left) and D0 (right)
experiments. Note that CDF is using the convention where $-2\beta_s=\phiS$.
  \protect\label{fig7}}
\end{figure*}
\begin{figure}[pht]
%\begin{minipage}{0.47\textwidth}
\includegraphics[width=3.0in]{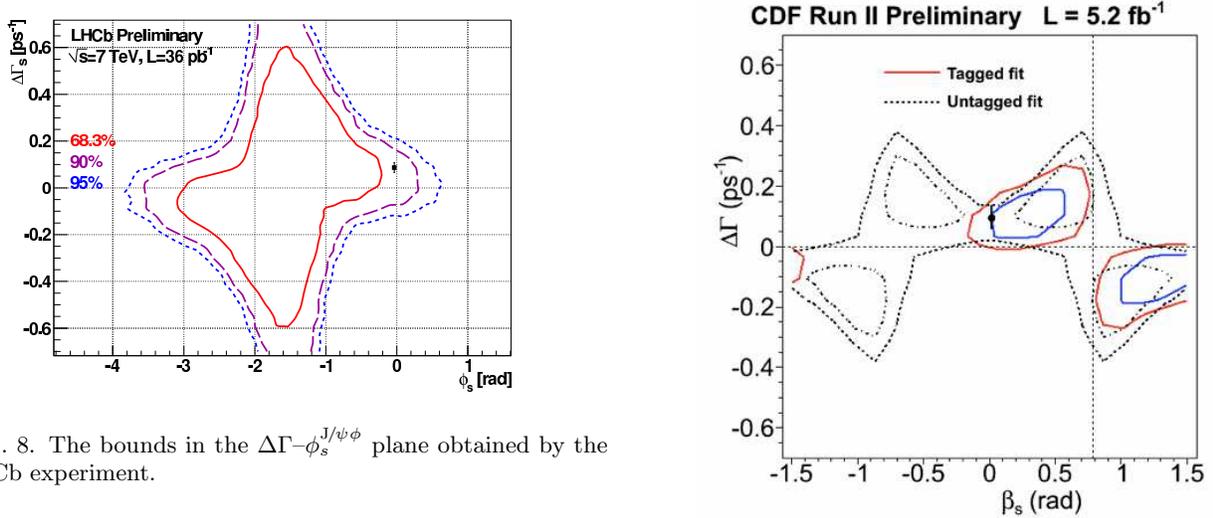}
%\centerline{\psfig{file=figs/Fig5,width=2.0in}}
%\vspace*{8pt}
\caption{The bounds in the $\Delta\Gamma$--$\phiS$ plane obtained by the LHCb experiment.
 \protect\label{fig8}}
\end{figure}%
%\end{minipage}
%\hfill
%\begin{minipage}{0.47\textwidth}
\begin{figure}[pht]
\includegraphics[width=3.0in]{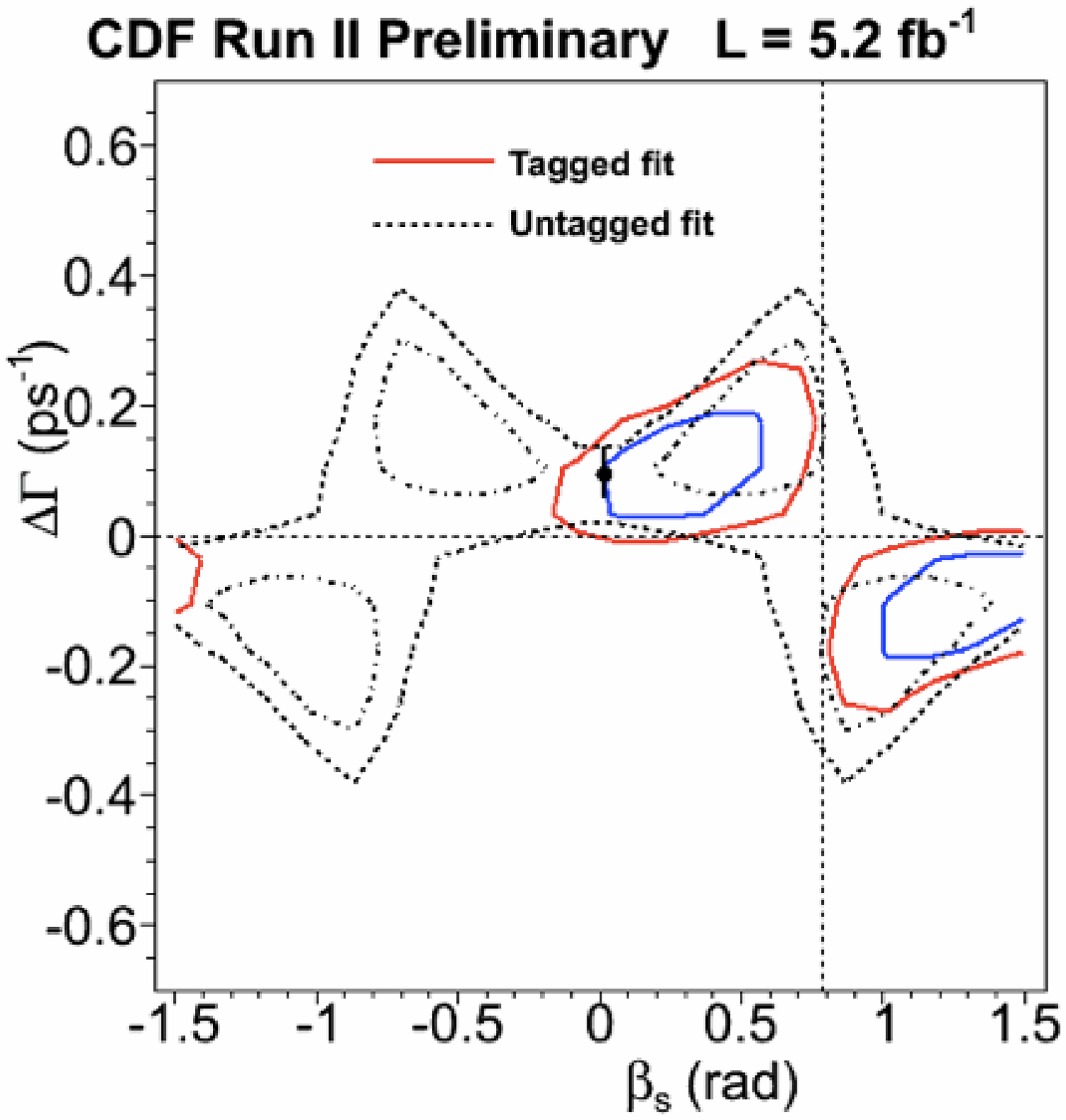}
%\centerline{\psfig{file=figs/overlay_fullyadjusted,width=2.0in}}
%\vspace*{8pt}
\caption{The comparison of bounds in $\Delta\Gamma$--$\beta_s$ plane obtained by the flavour-tagged (full
lines) and the flavour-untagged (dashed lines) analysis from the CDF experiment.
 \protect\label{fig9}}
%\end{minipage}
\end{figure}%
In Fig.~\ref{fig7} we show the confidence level contours in the $\Delta\Gamma$--$\phiS$ plane from the CDF 
and D0 experiments and in Fig.~\ref{fig8}, we show the contours derived by the LHCb experiment. The reader 
should be aware that CDF uses a different convention than D0 and LHCb with $-2\beta_s=\phiS$. As the
standard model is a special case, each experiment derives consistency between the data and the standard 
model. The consistency is characterized by the p-value, which is 44\% at CDF corresponding to about $0.8$ 
standard deviations and LHCb finds a p-value of 22\% which corresponds to about 1.2 standard deviations. 
Those tests provide an answer to the question of whether both $\Delta\Gamma$ and \phiS are simultaneously
consistent with the standard model. D0 does not evaluate an answer to this question, but from
Fig.~\ref{fig7} we can see that the agreement is equivalent to little more than one
standard deviation. It is worth noting that while CDF and LHCb allow for two solutions and
therefore have bounds which are symmetric, D0, by constraining the strong phases, allows only for one
solution. The fact that there are still two solutions in their result is an artefact of additional
approximate symmetries of the problem. Some difference between those approximately symmetric
solutions is seen, but the statistics are not sufficient to decide between them.
Before moving on, we come back to the point of deriving constraints without
flavour tagging. In Fig.~\ref{fig9} we show constraints from the untagged analysis of the CDF experiments
together with the result from Fig.~\ref{fig7}. As one can see the size of the contours is not very
different between the two analyses and the main help from the flavour tagging is in removing two out of 
the four solutions.
The slight shift of the two results is due to the difference in the importance of each candidate for two
analyses. The untagged analysis was performed by LHCb but no useful constraint could be derived with
the current statistics \cite{LHCb:untagged}. D0 also performed a fit without flavour tagging and obtained 
a result consistent with the tagged fit.

What is more interesting for some people is simply the value of the \CP-violating phase \phiS, rather than 
the allowed region in two-dimensional space. To obtain this, CDF and LHCb basically repeat the procedure used 
for the two-dimensional case where $\Delta\Gamma$ is also treated as an unimportant parameter and
maximize likelihood over it. The procedure yields $\phi_s^{\Jpsi\phi} \in [-3.1,-2.16]\cup[-1.04,-0.04]$ at 68\%
confidence level and $\phi_s^{\Jpsi\phi} \in [-\pi,-1.78]\cup[-1.36,0.26]\cup[2.88,\pi]$ at 95\%
confidence level at the CDF experiment. At LHCb the allowed regions are $\phi_s^{\Jpsi\phi} \in
[-2.7,-0.5]$ at 68\% confidence level and $\phi_s^{\Jpsi\phi} \in [-3.5,0.2]$ at 95\% confidence
level. The D0 experiment does not provide a result in this way, but their values can
be translated into approximate one-dimensional results as the correlation between $\Delta\Gamma$ and
$\phi_s^{\Jpsi\phi}$ is reasonably small, about $-18\%$. The intervals they obtain are
$\phi_s^{\Jpsi\phi} \in [-1.12, -0.38]$ at 68\% confidence level and $\phi_s^{\Jpsi\phi} \in
[-1.65,0.24]\cup[1.14,2.93]$ at 95\% confidence level without taking into account any correction for 
non-Gaussian behaviour. Also in the case of one-dimensional tests, there is a reasonable agreement between 
the standard model and data. 

As the CDF analysis implements also an s-wave contribution, it was possible to check also how large an 
effect it introduces. 
\begin{figure*}[pht]
\includegraphics[width=3.0in]{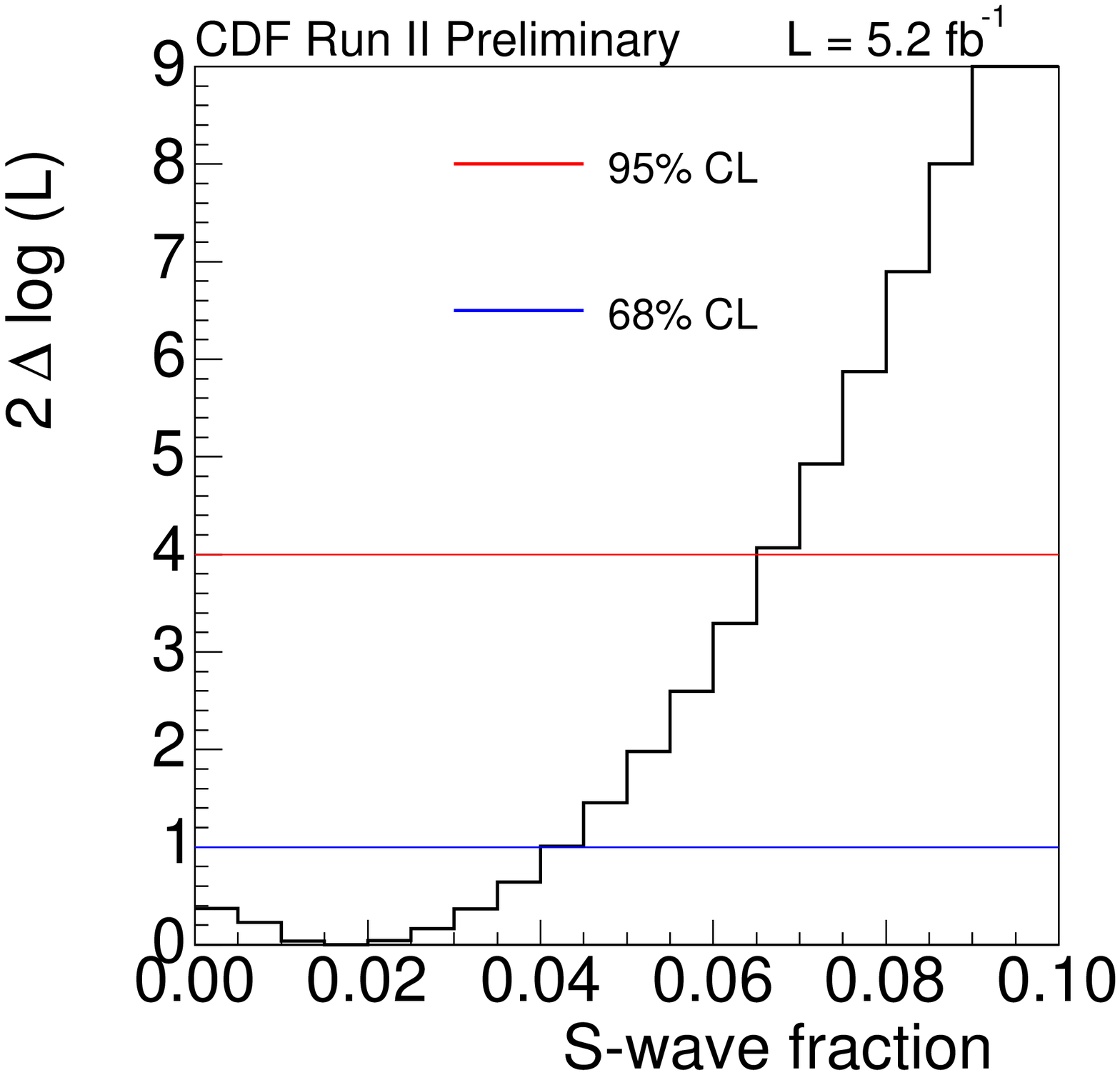}
\includegraphics[width=3.0in]{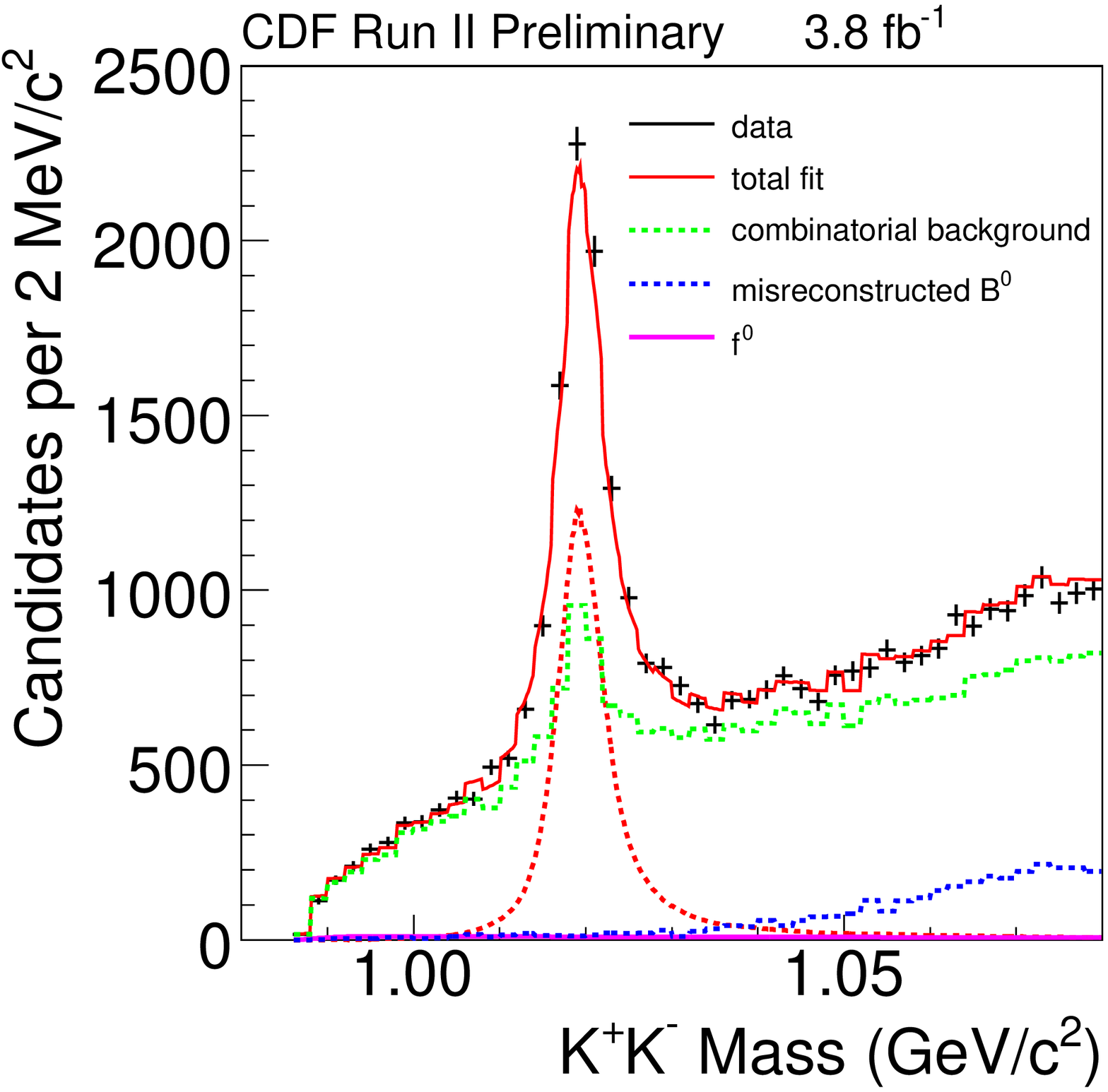}
%\centerline{\psfig{file=figs/swavefrac-lp,width=2.0in}\hspace{1cm}
%            \psfig{file=figs/KK_mass_1p8_f0,width=2.0in}}
%\vspace*{8pt}
\caption{The likelihood profile over the s-wave fraction from the full angular fit at CDF (left). The
invariant mass of the kaon pair together with the fit from the CDF experiment (right). The fit itself 
includes the s-wave with fraction fixed from the full angular fit.
  \protect\label{fig10}}
\end{figure*}
First, in Fig.~\ref{fig10} we show the likelihood profile for the amount of s-wave contribution and
the invariant
mass of the kaon pair, which is not used in the fit. From the full angular fit the obtained s-wave
fraction is consistent with zero. From the likelihood profile one can set a Bayesian upper limit of
about 7\% at 95\% credibility level for the s-wave fraction within the selected sample. 
One should note that this fraction is
selection-dependent. As a cross-check one can also check whether the kaon pair invariant mass is
consistent with the composition found in the full angular fit. We show the distribution together
with the fit where the s-wave fraction is fixed to the best value from the angular fit in Fig.~\ref{fig10}. 
As one can see, the model fits the data well and therefore adds additional confidence into the treatment 
adapted by the CDF experiment. Finally also a check for the effect of the s-wave can be made. In this CDF 
compared likelihood contours between fits with the s-wave allowed to float, and a fit with the s-wave 
fraction fixed to zero, and there is almost no visible difference between them. 
\begin{figure*}[pht]
\includegraphics[width=3.0in]{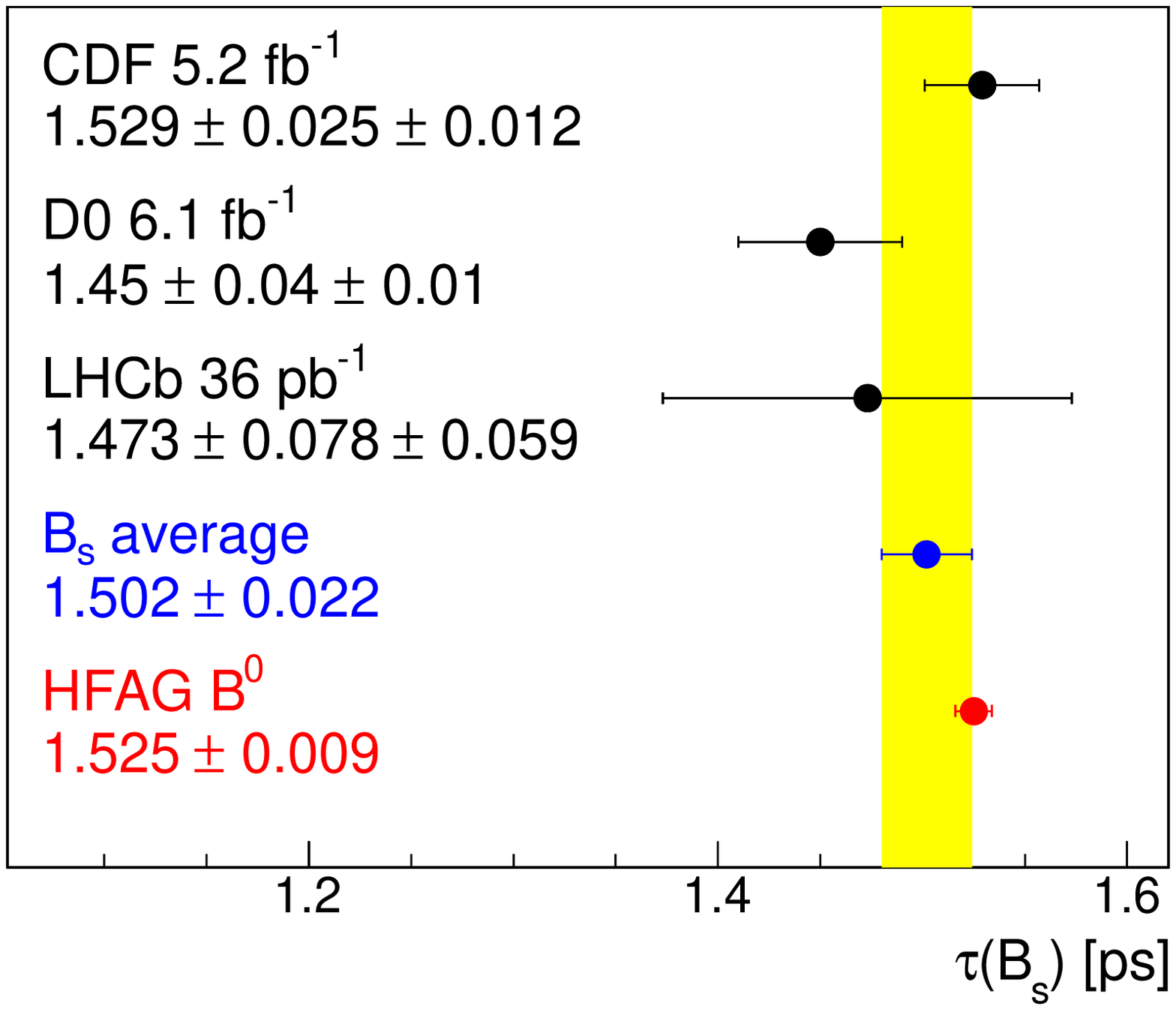}
\includegraphics[width=3.0in]{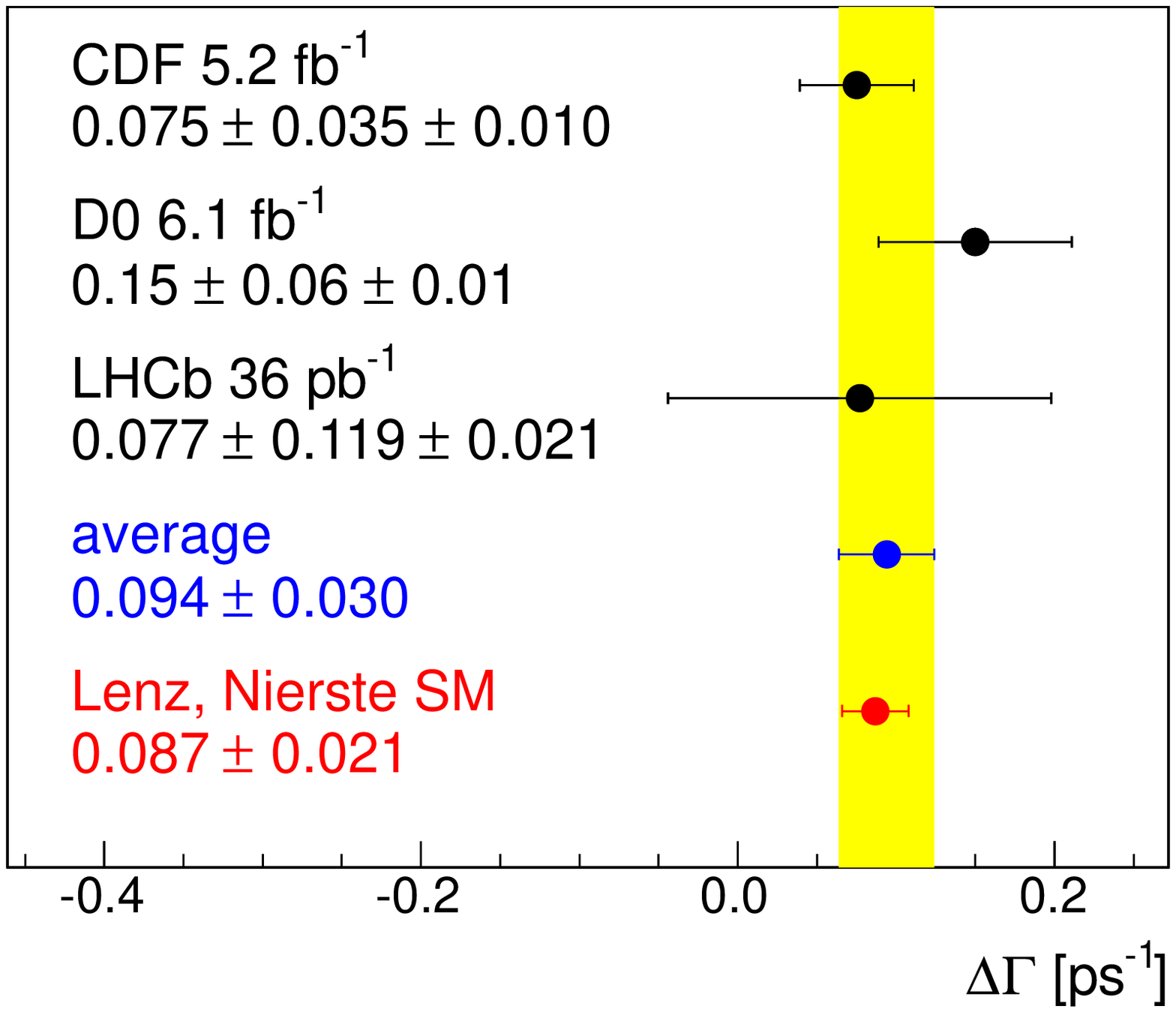}
%\centerline{\psfig{file=figs/compareLifetime,width=2.0in}\hspace{1cm}
%            \psfig{file=figs/compareDG,width=2.0in}}
%\vspace*{8pt}
\caption{The comparison of the measured mean lifetimes together with the $B^0$ lifetime which is predicted to 
be the same with high accuracy (left) and the decay width difference $\Delta\Gamma$ (right). %The point at
%bottom for lifetime is $B^0$ from Ref.~\cite{Asner:2010qj} and for $\Delta\Gamma$ is theoretical preduction
%from Ref.~\cite{Lenz:2011ti}.
  \protect\label{fig11}}
\end{figure*}
For completeness we show results for the lifetime and $\Delta\Gamma$ in Fig.~\ref{fig11}. They are
typically obtained in a fit which assumes no \CP violation and should therefore be treated in the
context of the standard model. For results on the amplitudes we kindly refer the reader to the original work
of the three experiments.

The important question is what the measurements tell us about the validity of the standard model and
potential new physics contributions. In all three experiments, the evaluation of the consistency with the 
standard model is available and in all three cases there is no significant departure from the
standard model. On the other hand, solutions obtained in all three experiments go in the same
direction from the standard model which suggests that there might be some effect of new physics.
While the combination would be interesting, unfortunately with the information publicly available it
is not possible to combine the results.
Moreover with the current precision it is not possible to exclude any \CP violation provided it lies
in the half-plane in which all the results are. Therefore practically any constraints on new physics
models from measurements presented here will be rather weak. This might not necessarily be the case when
combined with measurements of other quantities, but such a discussion is beyond the scope of this review.

\section{Prospects}
\label{sec:prospects}

% Comment about more data available and possible improvements. Should be pretty short.
What to expect in the near future? There are good prospects to see new results within a couple of months.
The Tevatron collider runs well and both CDF and D0 expect to collect about 10 \invfb of data by the end
of September 2011 when the Tevatron will terminate its operation. While some improvements would still be
possible, we do not expect a large gain beyond the increased statistics. In the meantime, LHC performance
is excellent with LHCb on track to collect about 1 \invfb of data by the end of this year. Given that
the first analysis was performed on only 37 \invpb this gives good prospects for a large statistical
improvement. Moreover as LHCb is an experiment which started to take data only last year it is
reasonable to expect some improvements which could help to constrain the \CP violation in \Bsboth mixing.
Finally while the ATLAS and CMS experiments did not present results in this area, measurement of the \CP
violation in \BsJpsiphi decay is in their plans with first results expected in the near
future. To conclude, in the near future the precision might be sufficient to see significant signal of large
\CP violation in \BsJpsiphi decay or constrain it to values close to the standard model. If it is
constrained close to the standard model, then question of the suppressed standard model
contributions will become important and could limit capability of bounding new physics until
progress on the understanding suppressed standard model contribution is made
\cite{Faller:2008gt,Ciuchini:2005mg,Ciuchini:2011kd}.

% In the final version I have to edit it to proper format,
% but for now I use bibtex to get right order.
%\bibliographystyle{mybib}
%\bibliographystyle{naturemag}
%\bibliography{ws-mpla}

\end{document}